\documentclass[paper=letter,12pt]{scrartcl}
\usepackage[version=4]{mhchem}
\usepackage[margin=1in]{geometry}
\usepackage{amsmath}
\usepackage{graphicx}
\usepackage{listings}
\usepackage{subcaption}
\usepackage{xcolor}
\usepackage{achemso}
\usepackage{float}
\usepackage{setspace}
\usepackage[margin=1in]{geometry}
\usepackage[english]{babel}
\usepackage[T1]{fontenc}
\usepackage{hyperref}
\hypersetup{
    colorlinks,
    linkcolor={black},
    citecolor={black},
    urlcolor={blue}
}

\title{
\usefont{OT1}{bch}{b}{n}

\fontsize{20}{25} \selectfont Pauli exclusion as a necessary condition for the existence of the \ce{H2 + H^.} atom transfer barrier\\[0.2in]
\fontsize{12}{8} \selectfont Shehan H. Fernando$^\text{a}$ and Alistair J. Sterling*$^\text{a}$\\[0.2in]
\fontsize{10}{15} \selectfont $^\text{a}$ Department of Chemistry and Biochemistry, University of Texas at Dallas, 800 W. Campbell road,  Richardson, Texas 75080-3021, United States\\[0.25in]
\fontsize{10}{15} \selectfont Email: Alistair.sterling@utdallas.edu
\date{}
}

\begin{document}
\maketitle


\sffamily{\textbf{The transition state barrier is a foundational concept in chemical kinetics, yet its physical origin is seldom investigated. In this study we analyze the barrier to hydrogen atom transfer in collinear \ce{H2}+\ce{H^.} on its Born--Oppenheimer surface. We extend upon the energy component analysis of chemical bonds to develop an electronic kinetic energy-based interpretation of the transition state barrier. This approach reveals that Pauli exclusion is necessary for the existence of the energy barrier to \ce{H2}+\ce{H^.} atom transfer, which is consistent with a complementary analysis using absolutely-localized molecular orbitals.}


\setstretch{1.5}
\rmfamily

The physical origin of the covalent chemical bond was a central enigma of quantum chemistry in its early days. This puzzle was solved by the Heitler--London model, which successfully explained the unexpectedly strong bond formed by two neutral \ce{H} atoms.\cite{heitlerlondon1927,frenking2000} The lowering of energy upon bond formation was interpreted in two ways: using the electronic kinetic energy (referred to as kinetic energy hereafter) by Hellmann and later by Ruedenberg,\cite{hellmann1933,ruedenberg1962} and the potential energy by Slater.\cite{slater1933} While these differing perspectives have yet to be fully reconciled, the past century of research through each lens has resulted in a rich understanding of the chemical bond.\cite{borkman1968,nordholm2020,levine2020,martin2022} 

Early efforts to extend chemical bonding arguments to explain chemical reactivity focused on the barrier-crossing reaction of the collinear \ce{H2} + \ce{H^.} triatomic system as an analogy to the \ce{H2} chemical bond.\cite{eyringpolanyi1931,slater1931} Such efforts culminated in several theories, including Fukui's frontier molecular orbital (FMO) theory,\cite{fukuiNOBEL} the avoided crossing formalisms by Longuet-Higgins and Abraham\cite{longuet1965} as well as Woodward and Hoffmann,\cite{woodward1969} and Shaik's work on avoided crossings combining ideas of MO theory and valence bond theory.\cite{shaik1981} Among these developments, the Klopmann--Salem and Fukui treatments of chemically interacting systems\cite{klopmann1968,salem1968,fukui1972,fukui1975} decomposed the minimum energy paths (MEP) of reacting systems into physically meaningful terms such as electrostatic and orbital interactions; such energy decomposition analysis (EDA) schemes (whose descendants include ALMO-EDA\cite{almoedaref_Khaliullin,almoedaref_horn} and EDA-NOCV\cite{morokuma1976,ORCAedanocv2025}) form the basis of many causal interpretations in modern chemistry.\cite{frenking2018}

To complement these EDA approaches, chemical bonding and reactivity have also been studied in terms of their energy components---kinetic ($T$) and potential ($V$) energies---in the spirit of Hellmann, Ruedenberg, and Slater.\cite{borkman1968,simons1974,nalewajski1993,tachikawa2000} Work by Nalewajski in 1980 used the component energies derived from the virial theorem 

\begin{align}
    T(\textbf{R}) &= -E(\textbf{R}) - \textbf{R}E'(\textbf{R}) \\
    V(\textbf{R}) &= 2E(\textbf{R}) + \textbf{R}E'(\textbf{R})
\end{align}

\noindent to validate potential energy surfaces constructed by various approximate methods available at the time, where $E$ is the Born--Oppenheimer (BO) energy, $\textbf{R}$ is the nuclear coordinate, and $E'$ are the Hellmann--Feynman forces.\cite{nalewajski1980} However, such an approach requires an exact wave function, and therefore provides approximate energy components for reaction analysis with approximate wave functions.\cite{lowdin1959}

Here we instead compute energy components directly from electronic wave functions to analyze collinear \ce{H2} + \ce{H^.} (Fig. 1). This approach enables us to also compute energy components of constrained wave functions to decouple competing physical mechanisms, thus establishing causality in transition state (TS) barrier formation. In particular, changes in kinetic energy can be interpreted in terms of electron localization and delocalization, signatures of competing Pauli exclusion effects and covalency. This reactivity analysis allows us to extend upon the recent work of Head-Gordon and coworkers who employed a natural orbital-based kinetic energy interpretation to investigate the roles of electron (de)localization during covalent chemical bond formation.\cite{sterling2024} 

\begin{figure}[H]
    \centering
    \includegraphics[width=0.6\columnwidth]{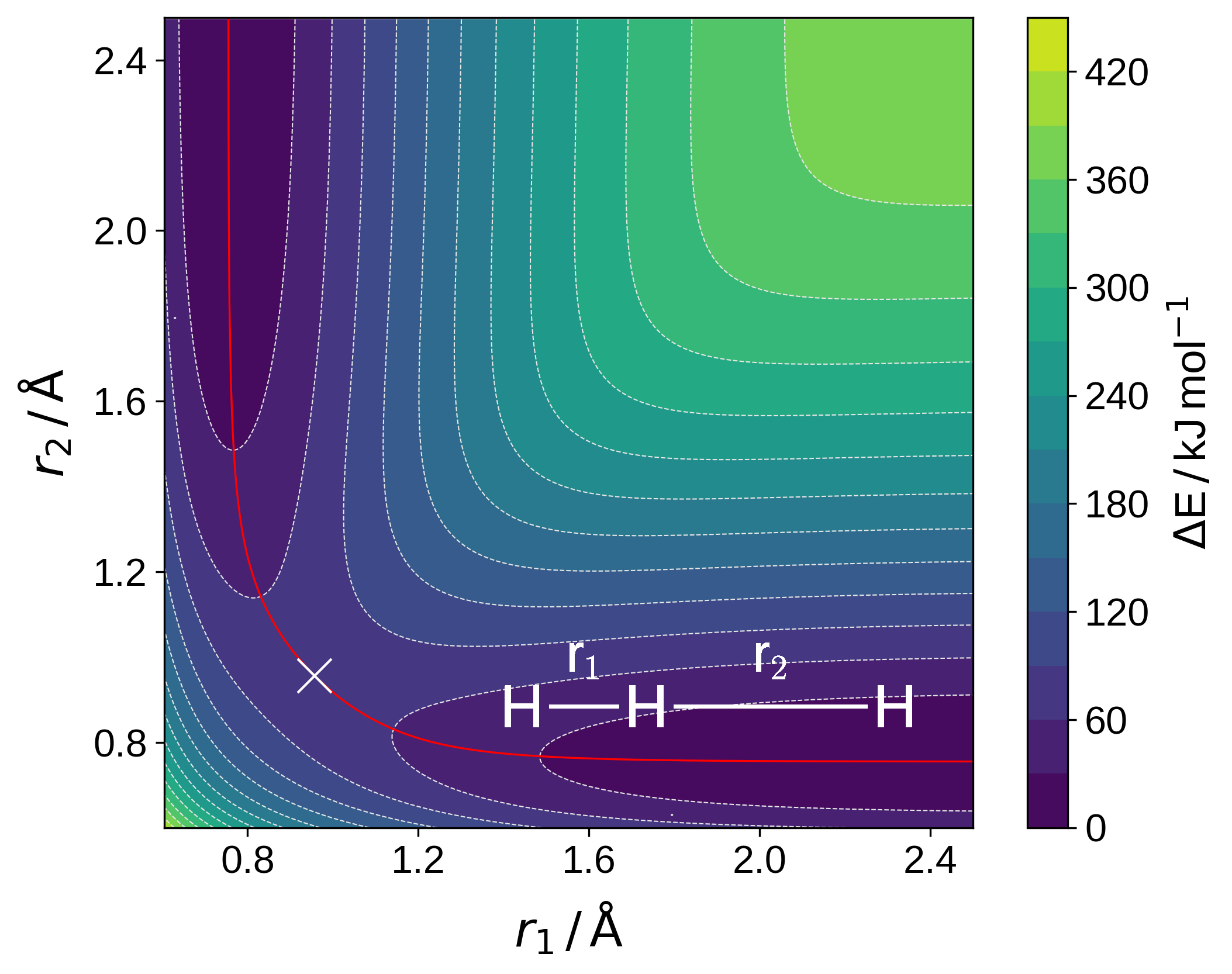}
    \caption{BO energy surface of the collinear \ce{H2} + \ce{H^.} atom transfer reaction (CASSCF(3,3)/aug-cc-pVTZ) with the trace (red) of minimum energy path crossing the transition state}
\label{doubletPES:main}
\end{figure}

Accordingly, kinetic energy is used here to interpret the physical interactions involved in the barrier-crossing hydrogen atom transfer (HAT) reaction of the \ce{H2} + \ce{H^{.}} system. This approach provides interpretation that complements the extensive prior work to model the BO energy surface of the \ce{H3} system with quantitative accuracy.\cite{truhlarwyatt1976,truhlarwyatt1977,liu1973,peterson2002,shaik1981} 

At the outset of our investigation, we hypothesized that, unlike in the parent \ce{H2} molecule, the pair of same-spin electrons in the \ce{H2} + \ce{H^{.}} system will cause additional effects due to Pauli exclusion. To examine the role of these effects on TS barrier formation, we envisaged ionizing the \ce{^2[H3]} system to produce a pair of states: singlet \ce{^1[H3]^{+}} in which only a pair of opposite-spin electrons remain, and triplet \ce{^3[H3]^{+}} in which only a pair of same-spin electrons remain. If Pauli exclusion is a necessary condition for the TS barrier in \ce{^2[H3]}, we should expect that the barrier should disappear in \ce{^1[H3]^{+}}, but remain in \ce{^3[H3]^{+}}. However, such an observation could be obscured by orbital relaxation effects upon ionization.

To decouple the effects of Pauli exclusion from orbital relaxation on the BO energy surface of the \ce{^1[H3]^{+}} and \ce{^3[H3]^{+}} systems, we adopted a strategy similar to the CASCI-based broken-bond orbital (BBO) analysis introduced by Head-Gordon.\cite{sterling2024} In the BBO approach, an intermediate wave function is generated by a CASCI calculation employing a set of CASSCF orbitals $\{\phi\}$ obtained at the broken-bond limit ($R_{BB} \gg R_\text{eq}$). Since SCF iterations are not performed and the CASCI step only solves the active space FCI problem without orbital optimization, the resulting intermediate state is completely free of orbital relaxation effects such as contraction and polarization. Here we adapt this approach to instead generate an intermediate singly-ionized state ($N \rightarrow N-1$, where $N$ is the number of electrons) using fixed CASSCF $\{\phi\}$---``pre-ionization natural orbitals'' (PINOs)---while also keeping the molecular geometry $\textbf{R}$ fixed. The energy change accompanying this ionization can be defined as follows

\begin{equation}
\begin{aligned}
    \Delta E(N \rightarrow N-1,\textbf{R})
    &= E_\text{CASSCF}(N-1,\textbf{R}) - E_\text{CASSCF}(N, \textbf{R}) \\
    &= \left( E_\text{CASSCF}(N-1,\textbf{R}) - E_\text{CASCI}(N-1, \textbf{R}) \right) \\
    &\quad + \left( E_\text{CASCI}(N-1, \textbf{R}) - E_\text{CASSCF}(N, \textbf{R}) \right) \\
    &\equiv \Delta E_\text{PINO}(N \rightarrow N-1,\textbf{R}) + \Delta E_\text{rlx}(N \rightarrow N-1,\textbf{R})
\end{aligned}
\end{equation}

The first term, $\Delta E_\text{PINO}(\textbf{R})$, is the ionization energy with fixed CASSCF $\{\phi\}$ (PINOs), while the second term, $\Delta E_\text{rlx}(\textbf{R})$, captures the orbital relaxation effects that result from ionization. The PINOs therefore completely isolate the effects of Pauli exclusion and electron delocalization from orbital relaxation on the BO surface. Both BBOs and PINOs can be considered subsets of non-optimized natural orbitals that employ reference orbitals from the broken-bond ($\textbf{R})$ limit or the unionized ($N$) limit, respectively. All calculations were performed using Q-Chem 6.2.1\cite{qchem6} with the aug-cc-pVTZ\cite{dunning1992} basis set.

Analysis of the kinetic energy component of the doublet (\ce{^2[H3]}) MEP, traced with the SciPy RBF interpolator,\cite{scipy2020} see SI for computational details) reveals two important features: (1) an initial increase in kinetic energy upon approach of $\ce{H^.}$ toward $\ce{H2}$, followed by (2) a subsequent kinetic energy decrease that reaches a minimum at the TS (Fig. \ref{NormTE:main} top, blue line). 

\begin{figure}[H]
\centering
	\includegraphics[width=0.6\columnwidth]{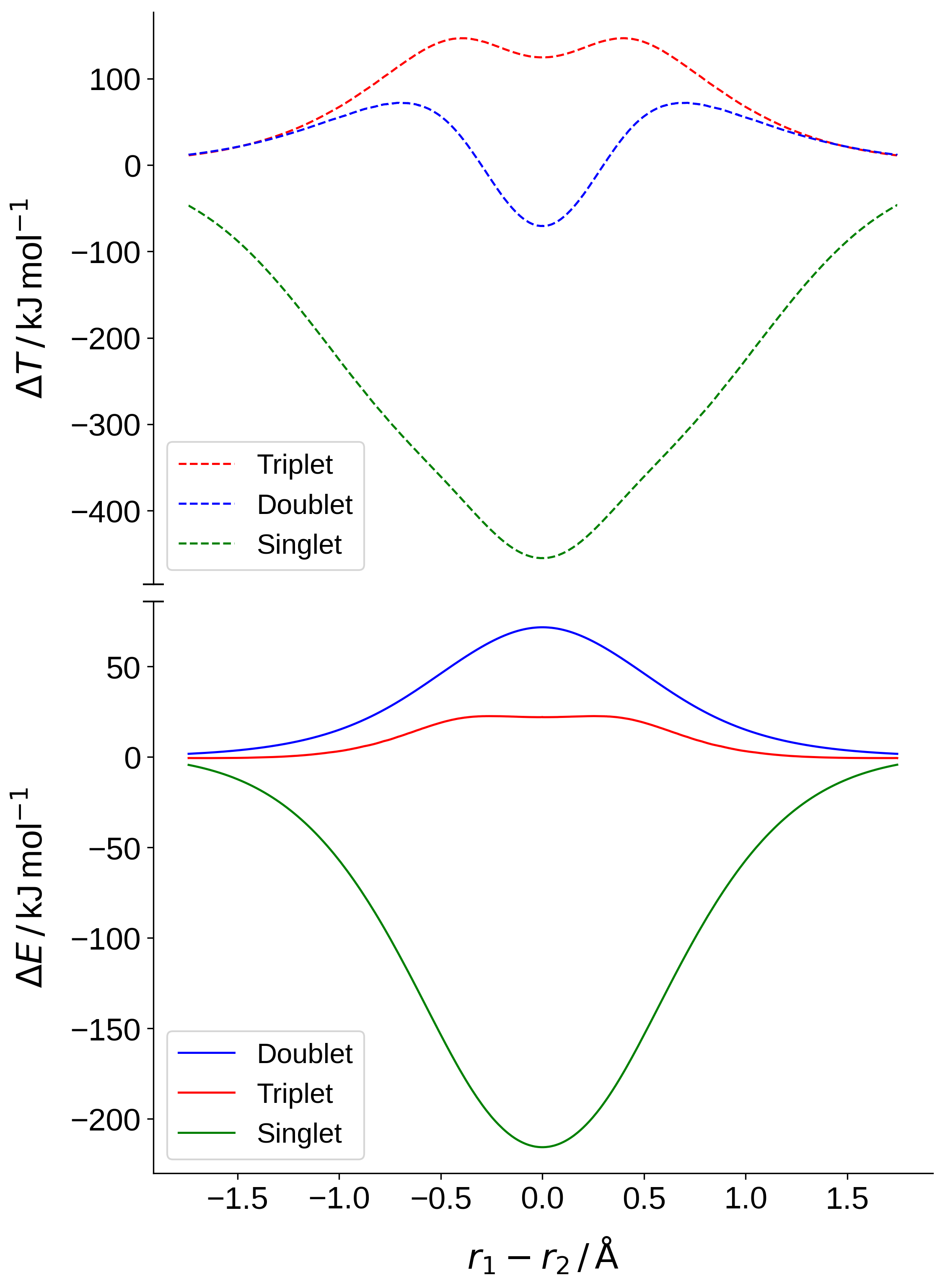}
	\caption{\textbf{(Top)} Total energy and \textbf{(Bottom)} kinetic energy of doublet ($\ce{^2[H3]}$, blue), triplet ($\ce{^3[H3]^{+}}$, red), and singlet ($\ce{^1[H3]^{+}}$, green) collinear \ce{H3} systems, normalized to the energy of the separated species on each respective surface. Singlet and triplet energies are computed using doublet PINOs on the doublet MEP.}
	\label{NormTE:main}
\end{figure}

Turning to PINO kinetic energy analysis, the triplet ($\ce{^3[H3]^{+}}$) surface shows an identical kinetic energy increase at long range compared to the doublet surface (Fig. \ref{NormTE:main} top, red line), while the singlet ($\ce{^1[H3]^{+}}$) surface displays the kinetic energy decrease typically associated with chemical bond formation.\cite{ruedenberg1962,sterling2024} Since the only difference between the triplet and singlet PINO surfaces is a single spin flip, we can attribute the initial kinetic energy increase solely to this spin flip, and thus to Pauli exclusion. Notably, long-range proportionality between $\Delta T$ and $\Delta E$ is required by the virial theorem for interactions decaying faster than $R^{-1}$ (see SI for derivation); in other words, if the overlap of electron densities of a pair of neutral molecules causes an increase in kinetic energy---for instance due to Pauli exclusion---then the Born--Oppenheimer energy must also increase.

The subsequent kinetic energy decrease can then be attributed to the delocalization of electrons across the three-atom system as it approaches the transition state. This electron delocalization effectively counteracts the localization effect of Pauli exclusion. 

The kinetic energy minimum that follows this decrease is required by the virial theorem and by symmetry, since the Hellmann--Feynman forces at a TS are zero, thus $\Delta E^\ddagger\,=\,-\Delta T^\ddagger$.\cite{nalewajski1980} In other words, a decrease in kinetic energy from reactants to the TS is a requirement for a TS barrier. The TS kinetic energy minimum in this symmetric transition state is consistent with the principle of minimum hardness of Toro-Labbé.\cite{toro1999} 

The competing localizing and delocalizing mechanisms that control kinetic energy changes in \ce{^2[H3]} can be quantified through analysis of natural orbitals ($\phi_i$) and their occupations ($\eta_i$) (Figure \ref{fig:orbitalKEoccSTCKEDprobamps}). The kinetic energy of the system is equal to the sum of the products of each one-electron NO kinetic energy and its respective occupation according to 

\begin{equation}
    T = \sum_i \eta_i \langle \phi_i \vert \hat{T} \vert \phi_i \rangle.
\end{equation}

We can therefore examine the behavior of each term in isolation to understand its contribution to the resultant kinetic energy of the whole system. $\phi_1$ captures electron delocalization between $\ce{H2}$ and $\ce{H^.}$ that decreases the kinetic energy (Fig. \ref{fig:orbitalKEoccSTCKEDprobamps}, (a) bottom row). This orbital remains almost doubly occupied ($\eta_{1} \approx 2$) throughout the reaction, such that the per-NO kinetic energy-lowering contribution is effectively doubled (Fig. \ref{fig:orbitalKEoccSTCKEDprobamps}, (b) top). Conversely, $\phi_2$, which is initially localized on the isolated $\ce{H^.}$ atom, develops nodal structure as it overlaps with the $\ce{H2}$ $\sigma$ orbital along the reaction coordinate due to node-induced electron confinement (Fig. \ref{fig:orbitalKEoccSTCKEDprobamps}, (b) middle row).\cite{kaupp2007,schwarz2020,sterling2024}

\begin{figure}[H]
    \centering
	\includegraphics[width=0.52\columnwidth]{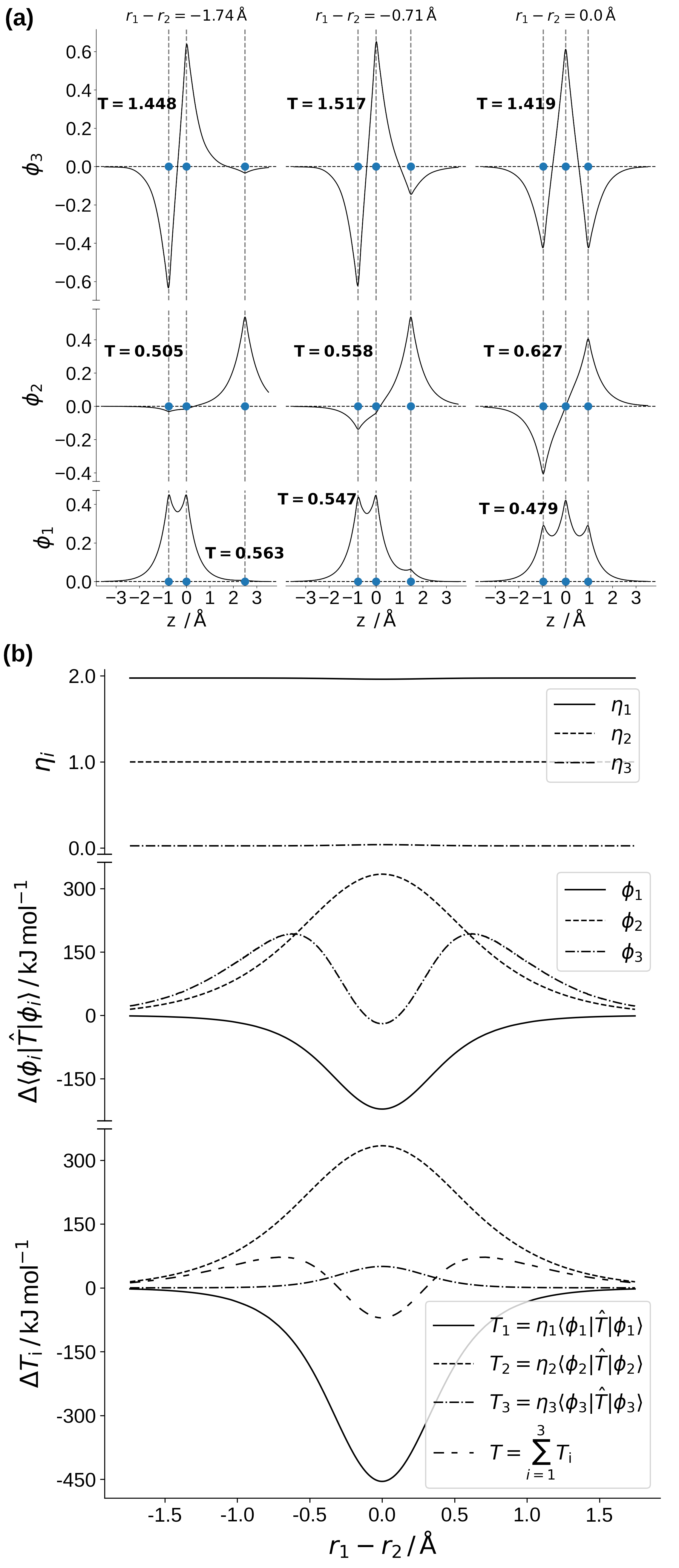}
	\caption{Probability amplitudes (processed using Multiwfn\cite{multiwfn1,multiwfn2}) and one-electron kinetic energies (in Ha) of the first three orbitals of the \ce{^2[H3]} system aligned along its z-axis \textbf{(a)}, Orbital occupancies ($\eta_i$, \textbf{(b) top}), one-electron kinetic energies ($\phi_i$, \textbf{(b) middle}) and the expectation values of orbital kinetic energy ($\eta_i\phi_i$) for the three active orbitals of \ce{^2[H3]} overlaid with expectation value of electronic kinetic energy (\textbf{(b) bottom}).}
	\label{fig:orbitalKEoccSTCKEDprobamps}
\end{figure}

 Pauli exclusion requires that the third electron occupies $\phi_2$, and the kinetic energy-raising effect that results from this singly-occupied $\phi_2$ consequently counteracts the kinetic energy-lowering effect of $\phi_1$. While the kinetic energy of $\phi_{3}$ increases toward the TS, before reaching a maximum and decreasing as the TS is reached, the effect of this kinetic energy is diminished by its negligible occupation ($\eta_3 \approx 0$) during the reaction (Fig. \ref{fig:orbitalKEoccSTCKEDprobamps}, (b) top). Thus, the kinetic energy changes of the \ce{^2[H3]} system are mainly derived from $\phi_1$ and $\phi_2$, which act to delocalize and localize the electrons, respectively. 

Fukui previously recognized the importance of the $\beta$ spin electron in the bond formation and breaking steps in the \ce{H2 + H^.} reaction, while the net $\alpha$ electron contribution is negligible.\cite{fukui1972} While not commented upon in their original work, their electron density calculations indicate that on approach of \ce{H^.} toward \ce{H2}, the $\beta$ electron from the \ce{H2} $\sigma$ bond can delocalize onto the \ce{H} atom---the mechanism captured by $\phi_1$ above. 

To validate the physical picture obtained from PINO composition and per-NO decomposition analyses, unrestricted ALMO-EDA calculations were performed on the doublet MEP at the $\omega$B97M-V/aug-cc-pVTZ level of theory\cite{wb97mv,dunning1992} using $\ce{H2}$
 and $\ce{H^.}$ as fragments.\cite{Head-Gordon_secgenEDA2016,almo-eda-unrest,Head-Gordon_secgenEDA2021} In this analysis, the interaction of $\ce{H2}$ with $\ce{H^.}$ is first divided into a geometric distortion ($\Delta E_\text{GD}$) and an intermolecular interaction term ($\Delta E_\text{INT}$); the latter term is further decomposed into frozen, polarization, and charge transfer interactions ($\Delta E_\text{INT} = \Delta E_\text{FRZ} + \Delta E_\text{POL} + \Delta E_\text{CT}$). The effects of Pauli exclusion-induced electron localization will be felt in $\Delta E_\text{FRZ}$ via the avoided overlap of occupied orbitals between fragments; this Pauli effect can be further isolated from permanent electrostatic and dispersion ($\Delta E_\text{FRZ} = \Delta E_\text{Pauli} + \Delta E_\text{ELEC} + \Delta E_\text{DISP}$).\cite{Head-Gordon2016} Interfragment electron delocalization resides solely within the $\Delta E_\text{CT}$ term. Orbital relaxation via electrical polarization is captured by the $\Delta E_\text{POL}$ term using dipole--quadrupole fragment electric-field response functions (nDQ model) that ensure a meaningful basis set limit.\cite{Head-Gordon2015}

The ALMO-EDA results show that $\Delta E_\text{FRZ}$ is the sole contributor to $\Delta E$ as the fragments begin to interact ($r_1 - r_2 > 1.5\,$\AA, Fig. \ref{fig:ALMOEDA} top). Within this frozen interaction, $\Delta E_\text{Pauli}$ is the only repulsive term in this region (Fig. \ref{fig:ALMOEDA} bottom), consistent with our conclusion that node-induced confinement in $\phi_2$ is responsible for electron localization (Fig. \ref{fig:orbitalKEoccSTCKEDprobamps}). As $r_1 - r_2$ decreases below $\approx 1.5\,$\AA, $\Delta E_{CT}$ begins to counteract Pauli repulsion; the result is a decrease in the kinetic energy gradient such that it first reaches an inflection point at $r_1 - r_2 = 1.2\,$\AA, then a maximum at $r_1 - r_2 = 0.8\,$\AA~(Fig. \ref{fig:ALMOEDA} top). This behavior is in accordance with the shallower onset of electron delocalization via $\phi_1$ relative to localization via $\phi_2$ (Fig. \ref{fig:orbitalKEoccSTCKEDprobamps}). As $\ce{H^.}$ continues to approach $\ce{H2}$, lengthening of the $\ce{H-H}$ bond softens Pauli repulsion while enabling increasing interfragment CT---bolstered by polarization of the softening electron density but costing $\Delta E_\text{GD}$---until $\Delta E_\text{INT}$ eventually becomes negative. Dominance of charge transfer, and to a lesser extent polarization, over Pauli repulsion in this bond-breaking/forming region concurs with the decrease in $\Delta T$ below 0. The physical origin of the \ce{H2 + H^.} barrier within the ALMO-EDA framework is therefore fully consistent with PINO and per-NO analysis, despite each being built upon a distinct theoretical foundation.

In summary, in this work we have shown that Pauli exclusion is a necessary condition for the transition state in the \ce{H2}+\ce{H^.} system. This finding is based on our newly-developed pre-ionization natural orbital analysis that isolates electron localization and delocalization from orbital relaxation, and is corroborated by an energy decomposition analysis using absolutely-localized molecular orbitals. Notably, electron delocalization at the TS inversely affects the barrier height; because $\Delta E^\ddagger=-\Delta T^\ddagger$, a large kinetic energy decrease means a higher barrier. Hence, a predictive relationship between the shape of the delocalization-dependent kinetic energy surface and the TS barrier height could be a useful tool to predict chemical reactivity from first principles. Future work will establish the generality of the conclusions put forward here to chemical reactions across the periodic table.

\begin{figure}[H]
	\centering
	\includegraphics[width=0.6\columnwidth]{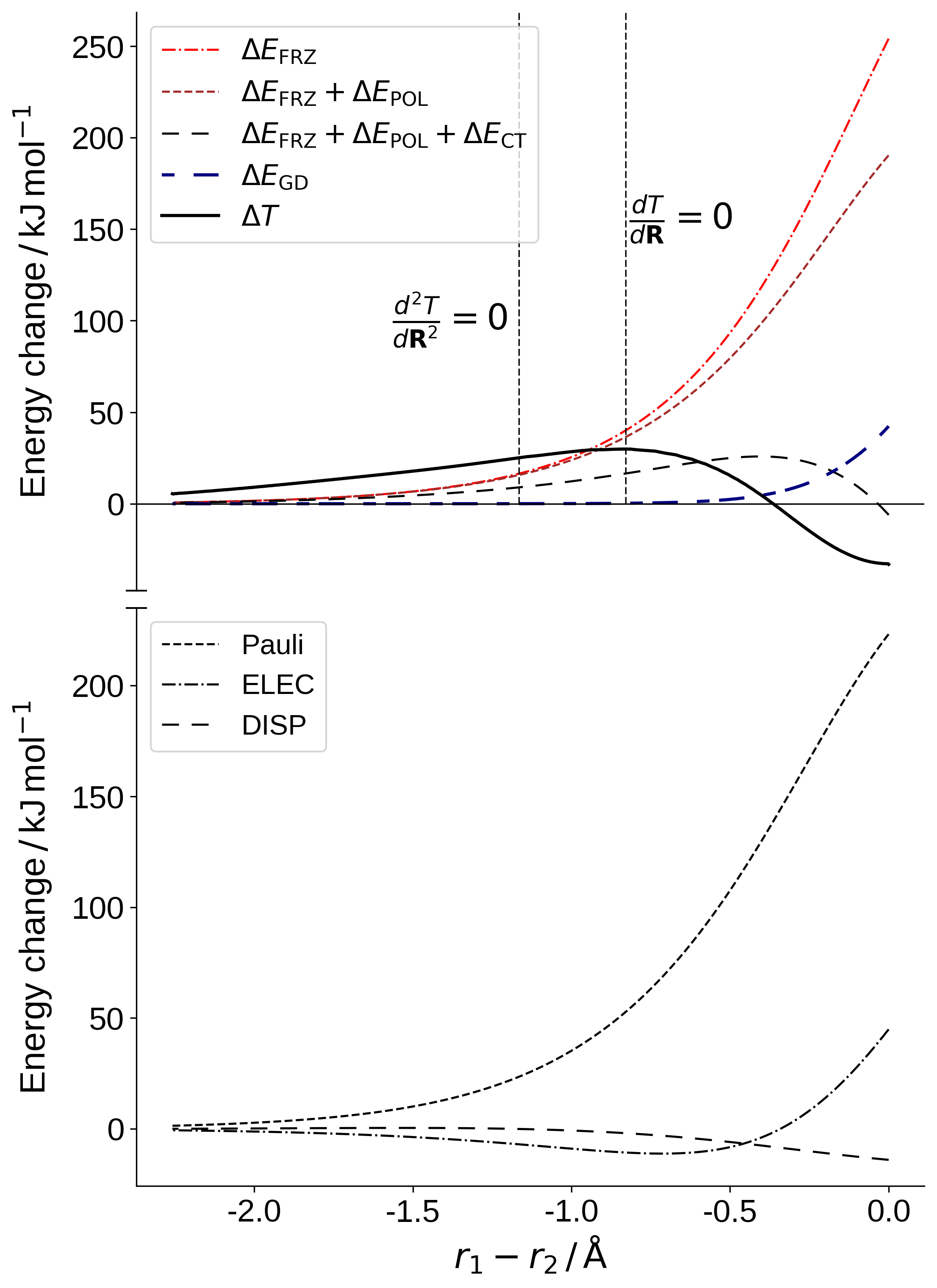}
	\caption{\textbf{(Top)} ALMO-EDA decomposition of $\Delta$E$_\text{int}$ into frozen (FRZ), polarization (POL) and charge transfer (CT) components, and the $\Delta$E$_\text{GD}$ and $\Delta T$ terms with inflection and turning points marked. \textbf{(Bottom)} Further decomposition of frozen energy term into Pauli, permanent electrostatic (ELEC) and dispersion (DISP) energy terms.}
	\label{fig:ALMOEDA}
\end{figure}

\section*{Author Contributions}
Fernando, S. H.: Conceptualization (equal), Methodology (equal), Software (lead), Validation (lead), Formal analysis (equal), Investigation (equal), Data curation (lead), Writing — original draft (lead), Writing — review \& editing (equal), Visualization (lead). Sterling, A. J.: Conceptualization (equal), Methodology (equal), Formal analysis (equal), Investigation (equal), Resources (lead), Writing — review \& editing (equal), Supervision (lead), Project administration (lead), Funding acquisition (lead).

\section*{Conflicts of interest}
There are no conflicts to declare.

\section*{Acknowledgements}
The authors thank Markus G. S. Weiss for helpful discussions, and acknowledge High-Performance Computing at The University of Texas at Dallas (HPC@UTD) for providing computing resources and support. This work was supported by the American Chemical Society Petroleum Research Fund (69496-DNI4), and A.J.S. acknowledges start-up support from The University of Texas at Dallas.

\section*{Data availability}

The data supporting this article have been included as part of the Supplementary Information. The code for analysis and plotting can be found at \url{https://github.com/SHFernando/Pauli-exclusion-in-H2-H-barrier}

\bibliography{refs.bib}

\end{document}


\title{
\linespread{2.5} \fontsize {18}{8} \selectfont Pauli exclusion as a necessary condition for the existence of the \ce{H2 + H^.} atom transfer barrier\\[0.2in]
\fontsize {14}{8} \selectfont Supporting information\\[0.25in]
\fontsize {12}{8} \selectfont Shehan H. Fernando$^a$ and Alistair J. Sterling$^{a*}$\\[0.25in]
\fontsize{10}{8} \selectfont $^\text{a}$ Department of Chemistry and Biochemistry, University of Texas at Dallas, 800 W. Campbell Road,  Richardson, Texas 75080-3021, United States\\[0.25in]
\fontsize{10}{8} \selectfont Email: alistair.sterling@utdallas.edu
\date{}
}

\maketitle
\setstretch{1.15}
\vspace{-0.3in}
\tableofcontents

\newpage

\section{Methods and calculation settings}

A summary of computational methods and implementations are given below. The default settings in Q-Chem 6.2.1 are used for calculations and any additional settings are mentioned in respective sections

\subsection{PES calculations}

PESs are calculated using CASSCF/aug-cc-pVTZ, DFT ($\omega$B97M-V/aug-cc-pVTZ) and CCSDT/aug-cc-pVTZ implementations in Q-Chem 6.2.1.\cite{qchem6,dunning1992} The surfaces are a collection of single-point energy calculations orchestrated by a Python 3.13.7 script that creates program input files and geometries within a given range of inter-nuclear distances. The total energy extracted from each calculation is used by the RBF interpolator implemented in SciPy\cite{scipy2020} to generate the trajectory of the MEP of the reaction.\\[0.2in]

The settings explicitly defined in the CASSCF input are,
\begin{itemize}
\ttfamily
\item SCF\_ALGORITHM              DIIS\_GDM
\item scf\_print = 1
\item point\_group\_symmetry = false
\item integral\_symmetry = false
\item unrestricted     True
\item INTERNAL\_STABILITY      True
\item INTERNAL\_STABILITY\_ITER 10
\item CAS\_SOLVER                 0 ! Required by CASCI for per orbital population analysis, used in CASSCF for consistency
\end{itemize}

The settings explicitly defined in the $\omega$B97M-V input are,
\begin{itemize}
\ttfamily
\item scf\_print = 1
\item SCF\_ALGORITHM              DIIS\_GDM
\item MAX\_SCF\_CYCLES             500
\item point\_group\_symmetry = false
\item integral\_symmetry = false
\item INTERNAL\_STABILITY      True
\item INTERNAL\_STABILITY\_ITER 25
\item unrestricted    true
\end{itemize}

The settings explicitly defined in the CCSDT input are,
\begin{itemize}
\ttfamily
\item mem\_total    12000
\item scf\_print = 1
\item SCF\_ALGORITHM              DIIS\_GDM
\item MAX\_SCF\_CYCLES             500
\item point\_group\_symmetry = false
\item integral\_symmetry = false
\item INTERNAL\_STABILITY      True
\item INTERNAL\_STABILITY\_ITER 25
\item unrestricted    true
\end{itemize}

\subsection{PINO calculations}

The PINO calculations are based on a CASSCF step followed by a CASCI step. Single point calculation input files are generated based on the trajectory geometries of MEP using a Python script.

\noindent The CASSCF first step is identical to input in the PES section.\\[0.1in]
The settings explicitly defined in the PINO CASCI second step input are,
\begin{itemize}
\ttfamily
\item CAS\_METHOD                 1
\item exchange hf
\item scf\_guess          read
 \item CAS\_N\_ROOTS                 1
\item CAS\_SOLVER                 0
\item gen\_scfman     True
\item max\_scf\_cycles   0
\item scf\_convergence  1
\item basis aug-cc-pVTZ
\item scf\_print = 1
\item point\_group\_symmetry = false
\item unrestricted    true
\item integral\_symmetry = false
\item SCF\_ALGORITHM  DIIS
\end{itemize}

\subsection{EDA (energy decomposition analysis) and non-perturbative COVP analysis based on $\omega$B97M-V/aug-cc-pVTZ}

Second generation EDA implementation of Q-Chem 6.2.1 is used. The polarization term is obtained in this version using the ``nDQ'' model which ensures a value with a meaningful basis-set limit.

\begin{itemize}
\ttfamily
\item scf\_print 1
\item point\_group\_symmetry = false
\item integral\_symmetry = false
\item unrestricted    true
\item INTERNAL\_STABILITY      True
\item INTERNAL\_STABILITY\_ITER 25
\item max\_scf\_cycles 250
\item EDA2                 1
\item scf\_algorithm DIIS\_GDM
\item internal\_stability\_davidson\_iter 500
\item gen\_scfman\_eda2  1
\item EDA\_VCT\_A  1
\item EDA\_POL\_A 1
\end{itemize}

\section{2D Born--Oppenheimer potential energy surfaces and saddle point location}

The minimum energy path (MEP) traced on the Born--Oppenheimer potential energy surface (PES) the collinear \ce{H2}+\ce{H^.} (\ce{^2H3} doublet) atom transfer is the backbone of our work. The CASSCF/aug-cc-pVTZ potential energy surface of this reaction was calculated by considering the whole valence space as active (CAS(3,3)). Tracing was done by the RBF interpolator function of SciPy;\cite{scipy2020} this trace is a smooth descent path compared to a path calculated using intrinsic reaction coordinate (IRC) method. The smoothness of the MEP is of high importance to electronic kinetic energy analysis (kinetic energy hereafter) as the kinetic energy is more sensitive to the fluctuations in the path than the total energy.\cite{slater1933,lowdin1959} 

\begin{table}[H]
\center
\begin{tabular}{ | P{1.3in} | P{1.3in} | P{1.3in} | P{1.3in} |}
\hline
Method &Asymptote energy / Ha & Saddle point energy / Ha & Barrier height / $^b$kJ mol$^{-1}$ \\
\hline
CASSCF (CAS 3,3) & -1.651423$^a$ & -1.624132 & 71.66 \\
\hline
$\omega$B97M-V & -1.655680$^a$ & -1.641870 & 36.26 \\
\hline
CCSDT & -1.672457$^a$ & -1.657014 & 40.55 \\
\hline
Reference (MRCI/aug-cc-pVTZ)\cite{peterson2002} & N/A & -1.657014 & $^c$40.55 \\
\hline
Reference (best estimate)\cite{peterson2002} & N/A & -1.659165 & $^d$40.20 \\
\hline
\multicolumn{4}{P{5in}}{\small $^a$ 100 \AA$~$separated \ce{H^.} and equilibrium \ce{H2} at given level of theory} \\
\multicolumn{4}{P{5in}}{\small $^b$ converted from Ha using factor=2625.49963947919 (kJ mol$^{-1})/$Ha} \\
\multicolumn{4}{P{5in}}{\small $^c$ asymptote given as separated equilibrium \ce{H2} and \ce{H^.}} \\
\multicolumn{4}{P{5.5in}}{\small $^d$ estimated as the median of lower and upper bounds of calculated barrier heights, see reference\cite{peterson2002} for details} \\
\end{tabular}
\end{table}

We compared the saddle point of the CASSCF surface with literature and it was noted the barrier height is overestimated.\cite{peterson2002} However, since we seek qualitative insight over quantitatively accurate energy surfaces, it is sufficient if the general features of the CASSCF surface are preserved with higher-level methods. The qualitative correctness of the CASSCF energy surfaces were verified using the CCSDT method. The calculations were performed using Q-Chem 6.2.1\cite{qchem6} and the aug-cc-pVTZ\cite{dunning1992} basis set. Since analytical kinetic energies are not available with CCSDT, these energy contributions were computed numerically (see Section 3 for details).

\begin{figure}[H]
\centering
\includegraphics[width=0.62\textwidth]{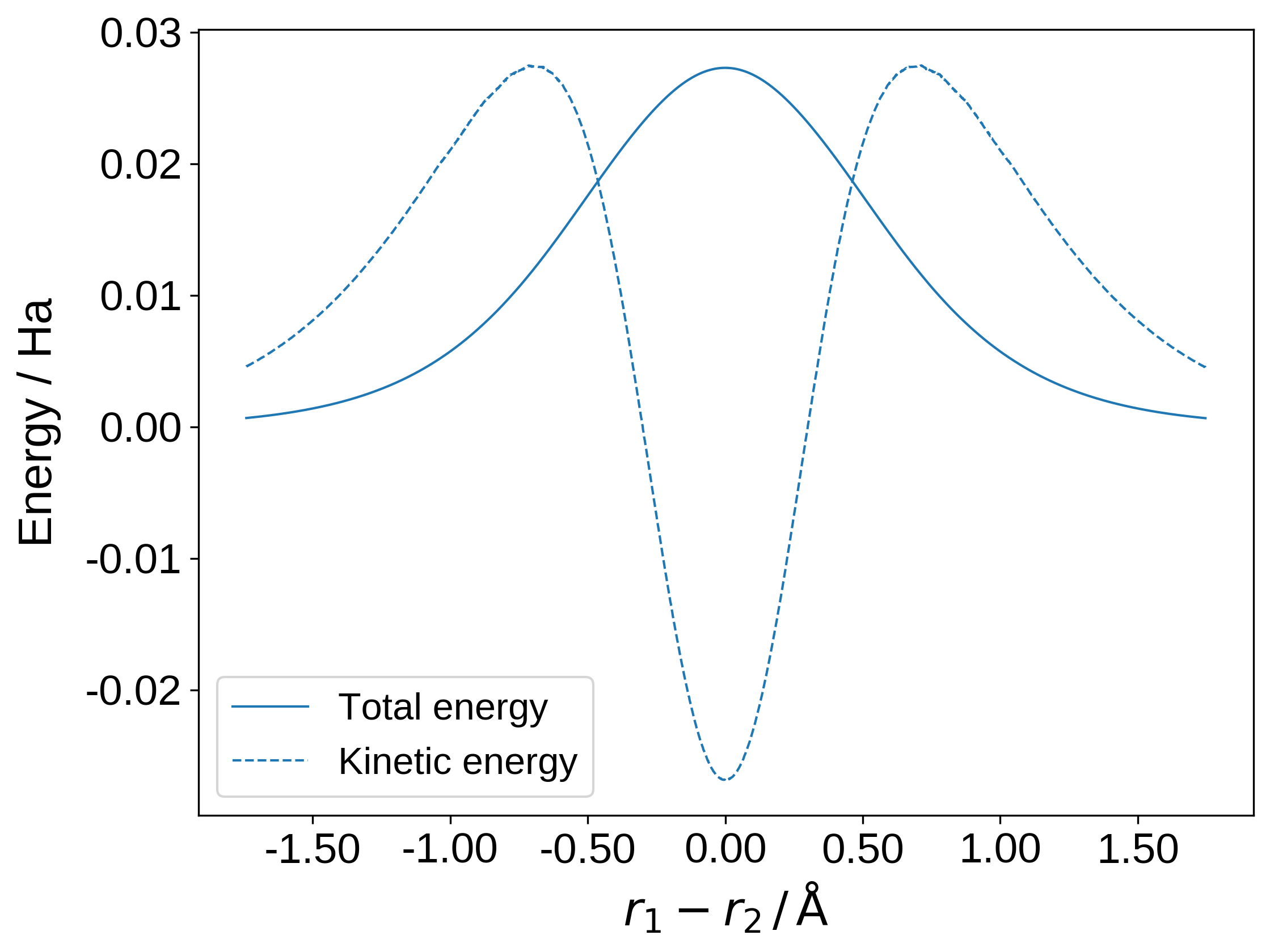}
\caption{The CASSCF kinetic energy composition of the collinear \ce{H2}+\ce{H^.} (\ce{^2H3} doublet) along the minimum energy path}
\end{figure}

\begin{centering}
\begin{figure}[H]
\centering
\includegraphics[width=0.7\textwidth]{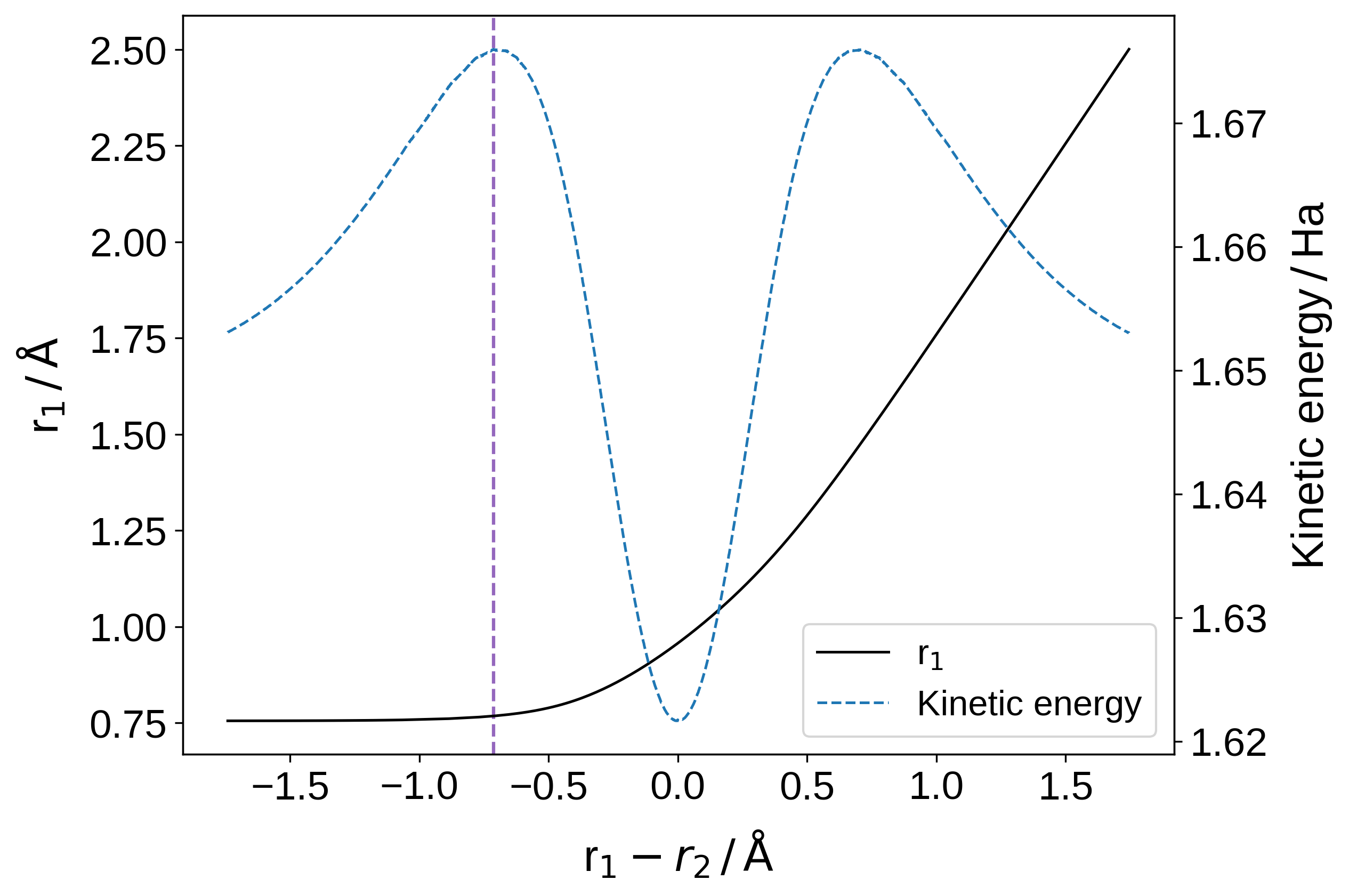}
\caption{Geometric distortion of the \ce{H2} bond length overlaid with the CASSCF kinetic energy of the \ce{^2H3}}
\end{figure}
\end{centering}

\begin{centering}
\begin{figure}[H]
\centering
\includegraphics[width=0.62\textwidth]{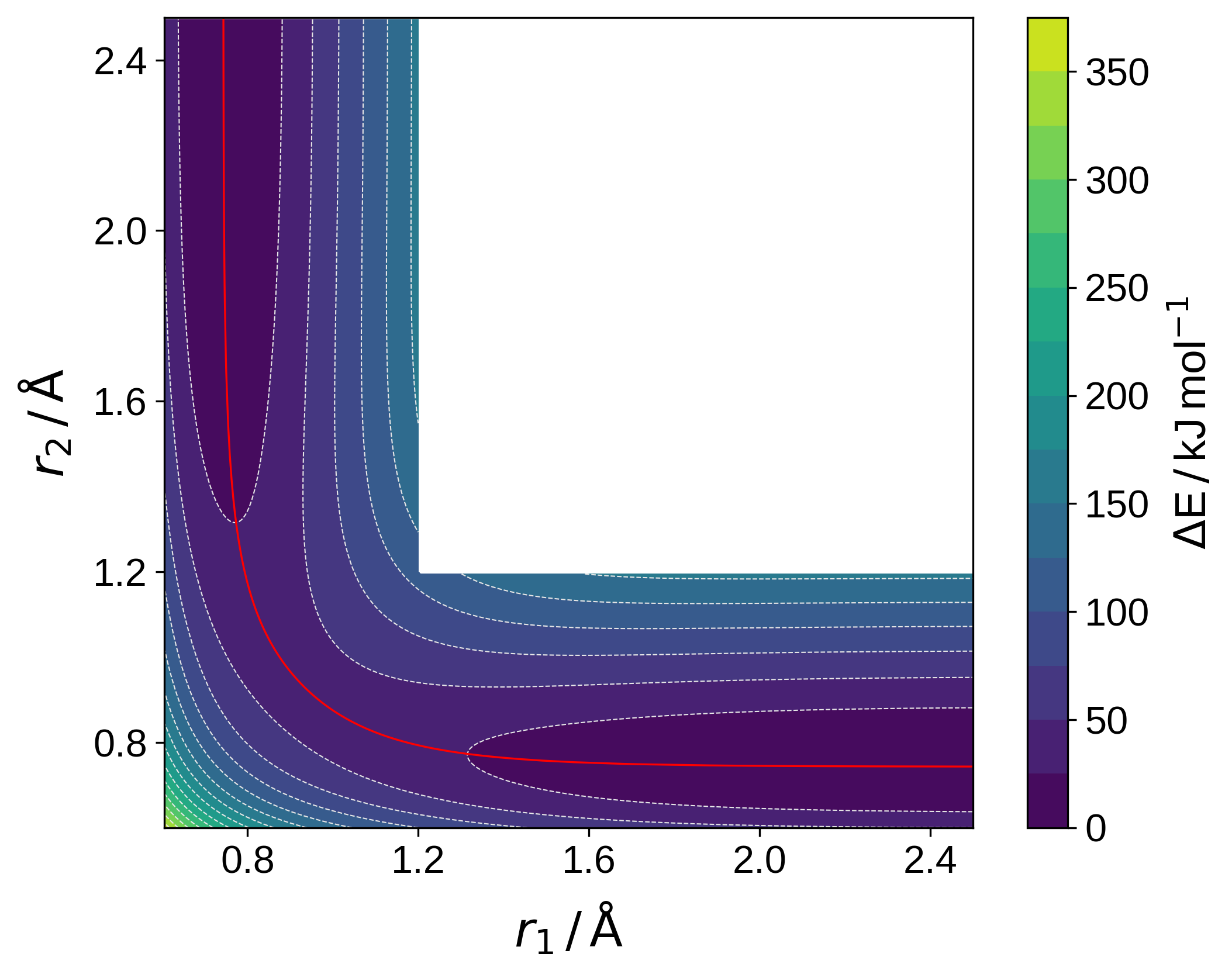}
\caption{The 2D potential energy surface of the collinear \ce{^2H3} at CCSDT/aug-cc-pVTZ}
\end{figure}
\end{centering}

\newpage

\section{Virial relations and asymptotic kinetic and potential energy topology}

Consider an exact, twice continuously differentiable Born–Oppenheimer energy curve for a nonrelativistic Coulomb system, with all other nuclear coordinates relaxed at each intermolecular separation \(R\). Let \(T\) denote electronic kinetic energy and \(V\) total Coulomb potential energy, including nuclear repulsion. All energy changes are relative to the separated fragments.

Let \(R_*\) be a stationary point, which may be a minimum or a transition state, with
\[
\Delta E'(R_*)=0,\qquad \Delta E(R_*)\ne0.
\]
Assume an inverse-power tail with the corresponding derivative behavior:
\[
\Delta E(R)\sim\frac{C}{R^n},
\qquad
\Delta E'(R)\sim-\frac{nC}{R^{n+1}},
\qquad C\ne0,
\qquad C \Delta E(R) > 0.
\]
These assumptions imply that both \(\Delta E\) and \(R\Delta E'\) tend to zero, and that \(\Delta E\) has the same sign at both \(R_*\) and at long range.

The molecular virial theorem, after subtracting the separated-fragment limit, gives
\[
\Delta T=-\Delta E-R\Delta E'.
\]
At the stationary point,
\[
\Delta T(R_*)=-\Delta E(R_*).
\]
At long range, substitution of the inverse-power tail gives
\[
\Delta T\sim(n-1)\Delta E.
\]
If we then assume \(n>1\) (\textit{i.e.}, the interaction is not purely governed by permanent electrostatics), \(\Delta T\) and \(\Delta E\) have the same sign sufficiently far out, while having opposite signs at \(R_*\), and \(\Delta T\to0\) as \(R\to\infty\). \(\Delta T\) therefore must change sign at finite separation greater than \(R_*\), and continuity thus guarantees at least one node in this range. 

Consequently, \(\Delta T\) has a negative stationary minimum if \(\Delta E(R_*)<0\), or a positive stationary maximum if \(\Delta E(R_*)>0\), at finite separation greater than \(R_*\).

For the potential energy curve, the virial theorem gives
\[
\Delta V=2\Delta E+R\Delta E'.
\]
Thus,
\[
\Delta V(R_*)=2\Delta E(R_*),
\qquad
\Delta V\sim(2-n)\Delta E.
\]
Therefore, if we assume \(n>2\), \(\Delta V\) has the same sign as \(\Delta E\) at \(R_*\), has the opposite sign sufficiently far out, and tends to zero as \(R\to\infty\). As above, continuity guarantees a node beyond \(R_*\), together with a positive stationary maximum if \(\Delta E(R_*)<0\), or a negative stationary minimum if \(\Delta E(R_*)>0\).

These conclusions establish the existence, though not uniqueness, of the nodes and extrema.

\section{Doublet (\ce{^2H3}) minimum energy path as singlet (\ce{^1H3+}) and triplet (\ce{^3H3+}): the effects of orbital relaxation}

A complete active space approach was used for the calculation of \ce{^1H3+} singlet and \ce{^3H3+} triplet. These singlet and triplet states were generated by removing an electron from (i.e., ionizing) the doublet (\ce{^2H3}) system along its MEP. The respective orbitals of the doublet MEP are passed onto a CASCI calculation without performing SCF iterations. Since the natural orbitals of the doublet system are used without optimization in the CASCI ionized states, we have introduced the term pre-ionization natural orbitals (PINOs). These orbitals and the broken bond orbitals (BBOs) of Head-Gordon\cite{sterling2024} can be generally regarded as non-optimized natural orbitals (NONOs) used in CASCI. The PINOs are orthogonal and non-singular transformations were not necessary before subsequent CASCI calculations.

\begin{figure}[H]
\centering
\begin{subfigure}[t]{0.5\textwidth}
\includegraphics[width=0.9\textwidth]{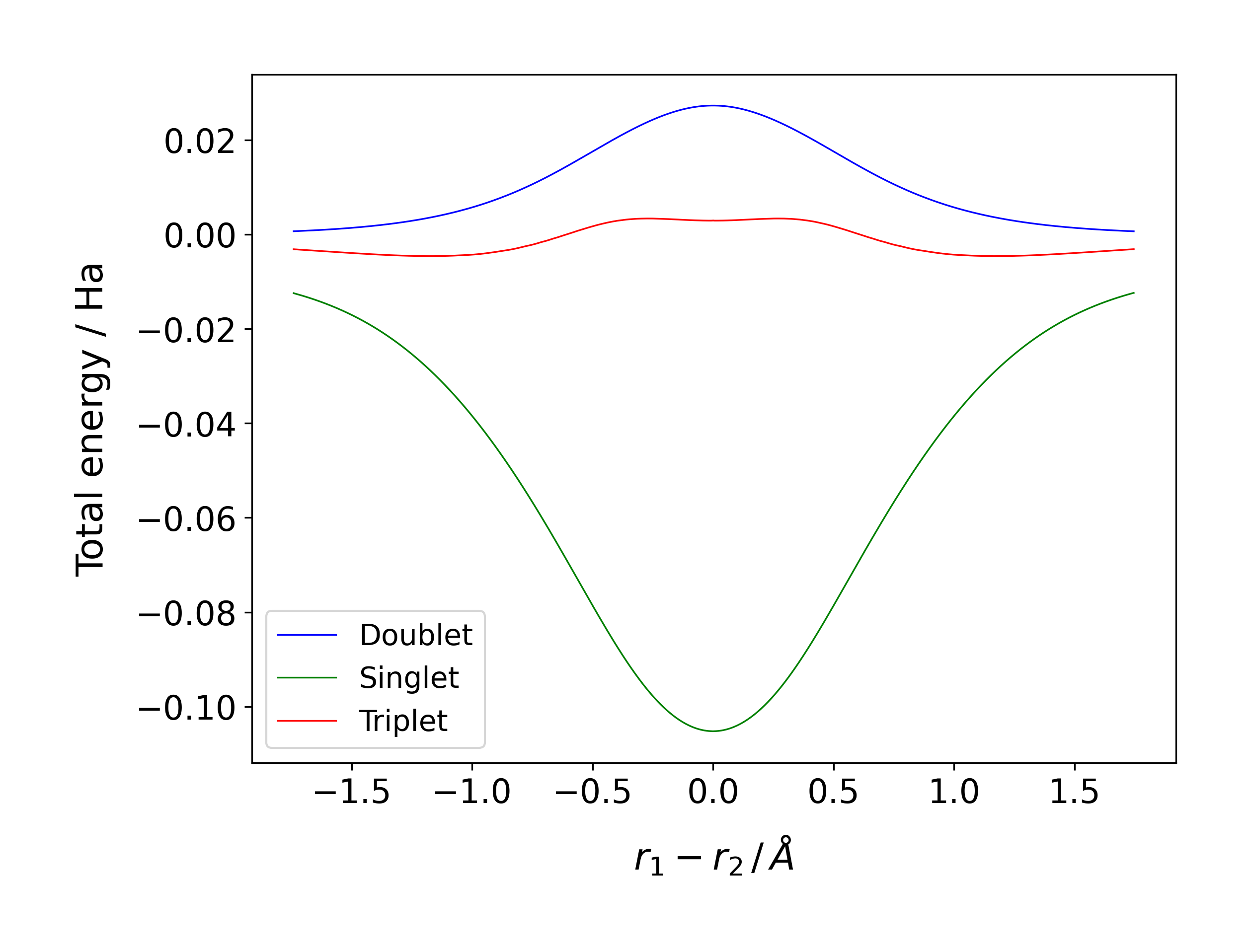}
\subcaption{}
\end{subfigure}%
\begin{subfigure}[t]{0.5\textwidth}
\includegraphics[width=0.9\textwidth]{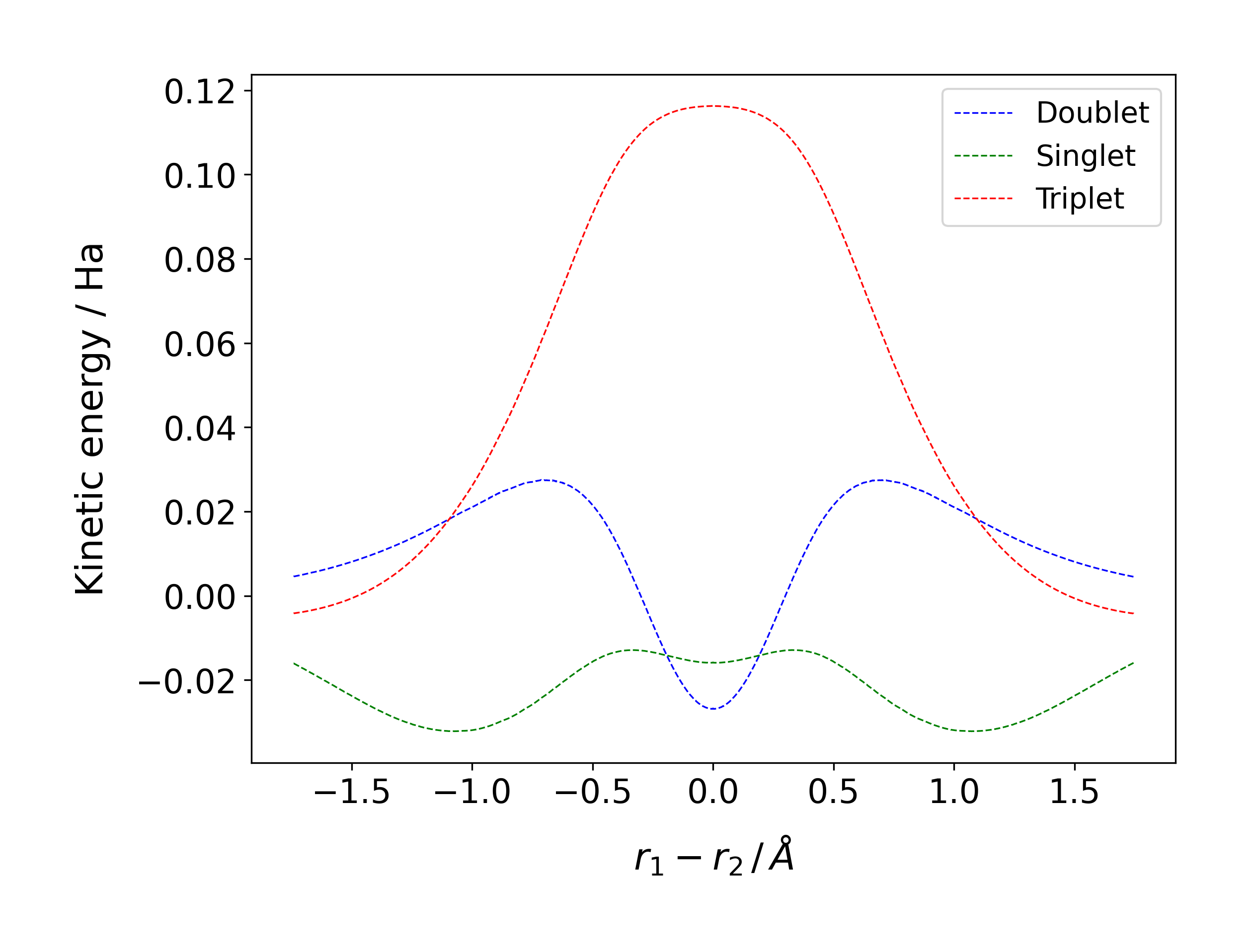}
\subcaption{}
\end{subfigure}
\caption{\textbf{(a)} The total energy and \textbf{(b)} the kinetic energy of the orbital relaxed (\textit{i.e.}, CASSCF) singlet (\ce{^1H3+}) and triplet (\ce{^3H3+}) states of the doublet (\ce{^2H3}) minimum energy path overlaid with the doublet (\ce{^2H3}) energy surfaces}
\end{figure}

\section{Doublet (\ce{^2H3}) kinetic energy derived from the virial theorem}

The kinetic energy of CCSDT method is not directly available. Thus, the virial theorem is used to calculate the kinetic energy from the energy gradients $\left( \frac{\partial E}{\partial R_{\alpha}}\right)$. 

\vspace{0.2in}

\noindent The virial theorem, where $R_{\alpha}$ is the internuclear distance between two nuclei, $T$ is the electronic kinetic energy ($\langle T\rangle$), $V$ is the potential energy ($\langle V\rangle$) and $E$ is the total energy can be written as,
\begin{equation}
2T + V + \sum_{\alpha=1}^NR_{\alpha}\frac{{\partial}E}{{\partial}R_{\alpha}} = 0
\end{equation}

since $T+V=E$

\begin{equation}
T = - \left( E + \sum_{\alpha=1}^NR_{\alpha}\frac{{\partial}E}{{\partial}R_{\alpha}}\right)
\end{equation}

CCSDT gradients (numerical or analytical) are not available from Q-Chem 6.2.1. Thus, CCSDT energy gradients are obtained numerically using the single point energies of the perturbed \ce{H3} system via the central differences method. 

\begin{figure}
\center
\begin{tikzpicture}
\node[anchor=center] at (0,0) {H};
\node[anchor=center] at (2,0) {H};
\node[anchor=center] at (5,0) {H};
\node[anchor=south] at (1,-0.1) {r$_{12}$};
\node[anchor=south] at (3.5,-0.1) {r$_{23}$};
\node[anchor=north] at (0,-0.1) {1};
\node[anchor=north] at (2,-0.1) {2};
\node[anchor=north] at (5,-0.1) {3};
\draw (0.2,0) -- (1.8,0);
\draw (2.2,0) -- (4.8,0);
\draw[<->,arrows = {Stealth[scale=1.5]-Stealth[scale=1.5]}] (-0.5,0.5) -- (0.5,0.5);
\node[anchor=south] at (0,0.6) {$\pm$ 0.001\AA};
\end{tikzpicture}
\caption{Perturbation of H atom 1 to get energy two of four required energy terms (E$_{+\,12}$ and E$_{-\,12}$). Similar perturbation of H atom 2 is also required at a given geometry to numerically calculate the energy gradient}
\label{fig:H3geompert}
\end{figure}
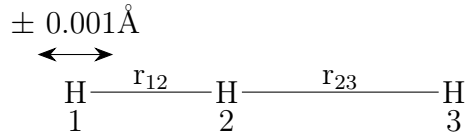

\begin{equation}
\frac{\partial E}{\partial r_{12}}=\frac{E_{-\,12}+E_{+\,12}}{0.002}
\end{equation}

Since the \ce{H3} system is aligned in the z-direction of Cartesian space, the perturbations are only required in the z-direction.(Fig:\ref{fig:H3geompert}) By choosing the central atom to be the Cartesian origin we can neglect forces on the central atom. Hence, the four perturbations of two outer atoms are considered in positive and negative z-direction for the gradient calculation. The gradient of kinetic energy is obtained using the central differences method and the noisy data is fitted with a spline (\texttt{make\_smoothing\_spline} of SciPy). Spline fitting agrees on points of interest in our kinetic energy analysis, such as maxima and minima, and we believe the qualitative information from the fitted splines are sufficient.

\begin{figure}[H]
\centering
\begin{subfigure}[t]{0.5\textwidth}
\includegraphics[width=0.9\textwidth]{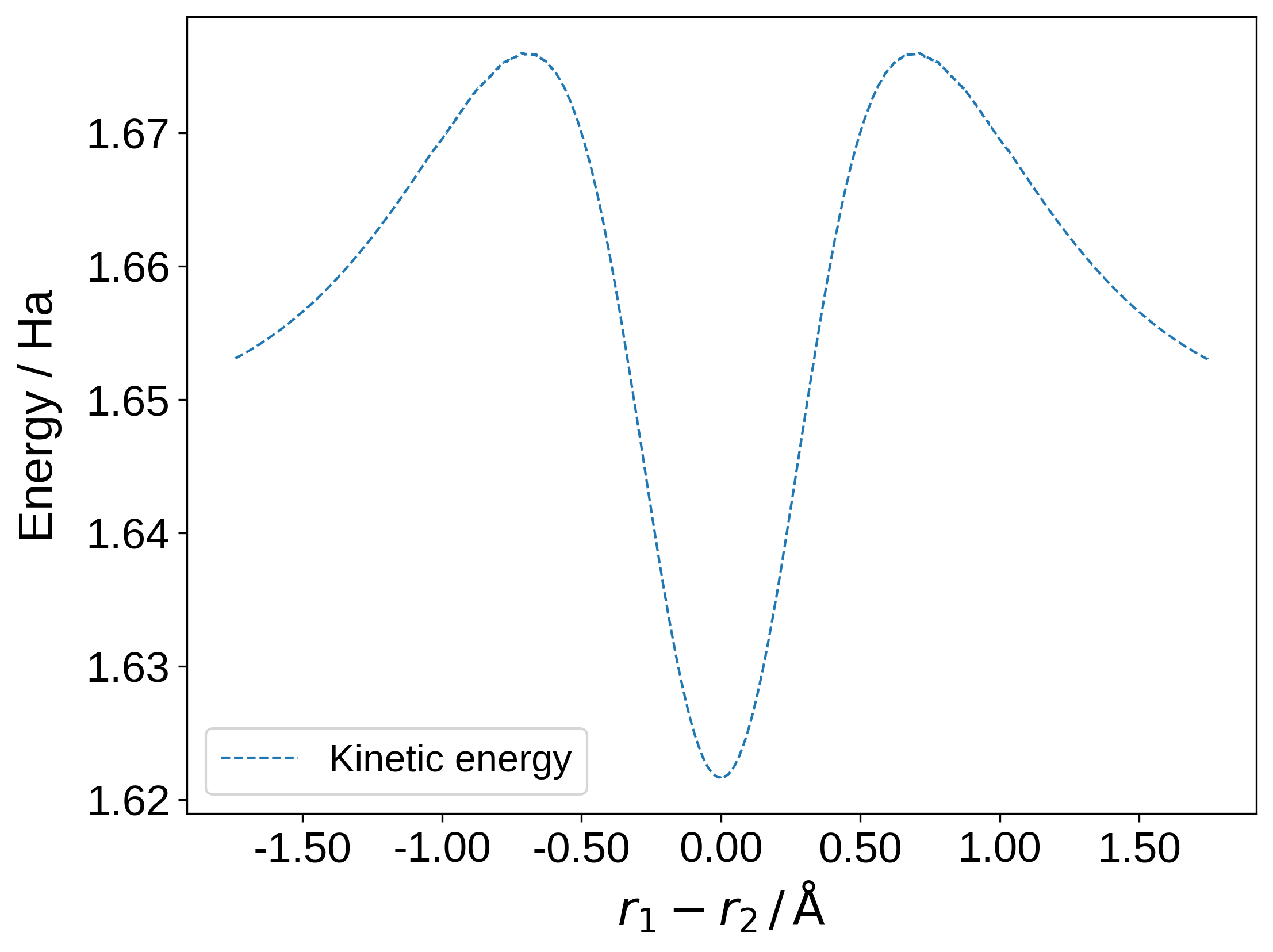}
\subcaption{}
\end{subfigure}%
\begin{subfigure}[t]{0.5\textwidth}
\includegraphics[width=0.9\textwidth]{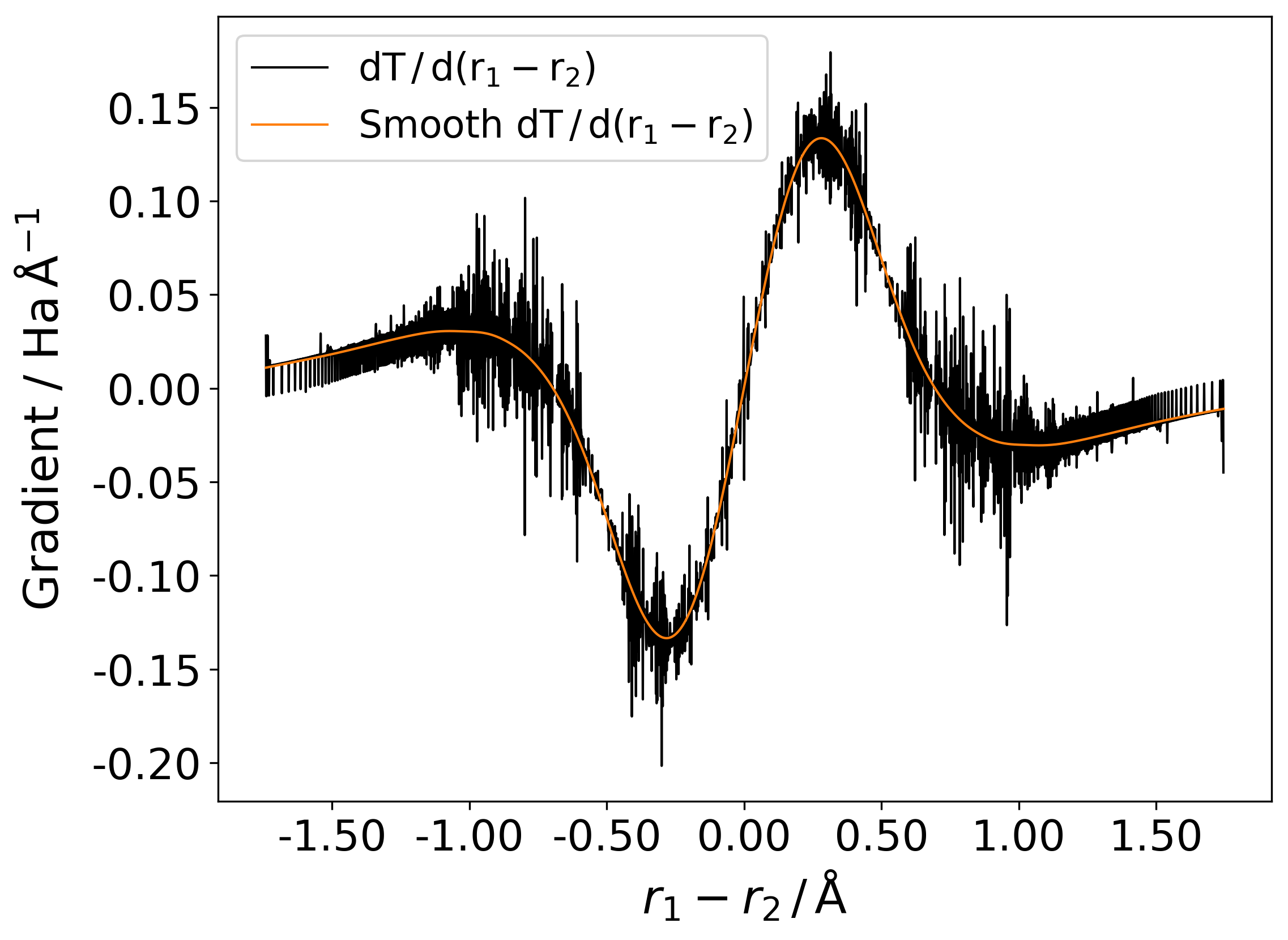}
\subcaption{}
\end{subfigure}
\caption{\textbf{(a)} CASSCF/aug-cc-pVTZ  kinetic energy calculated from the wavefunction, \textbf{(b)} gradient of the kinetic energy and the fitted spline}
\end{figure}

\begin{figure}[H]
\centering
\begin{subfigure}[t]{0.5\textwidth}
\includegraphics[width=0.9\textwidth]{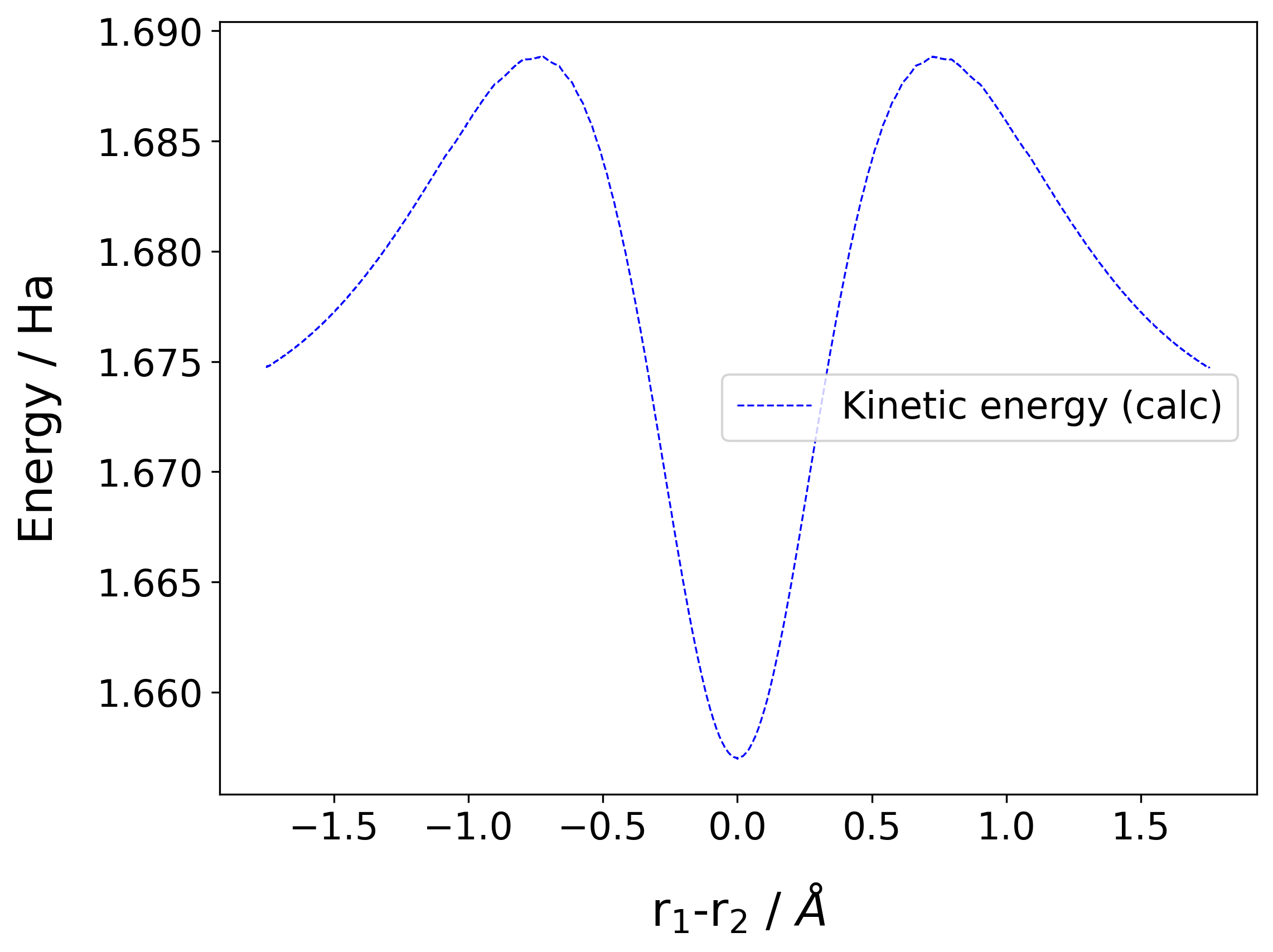}
\subcaption{}
\end{subfigure}%
\begin{subfigure}[t]{0.5\textwidth}
\includegraphics[width=0.9\textwidth]{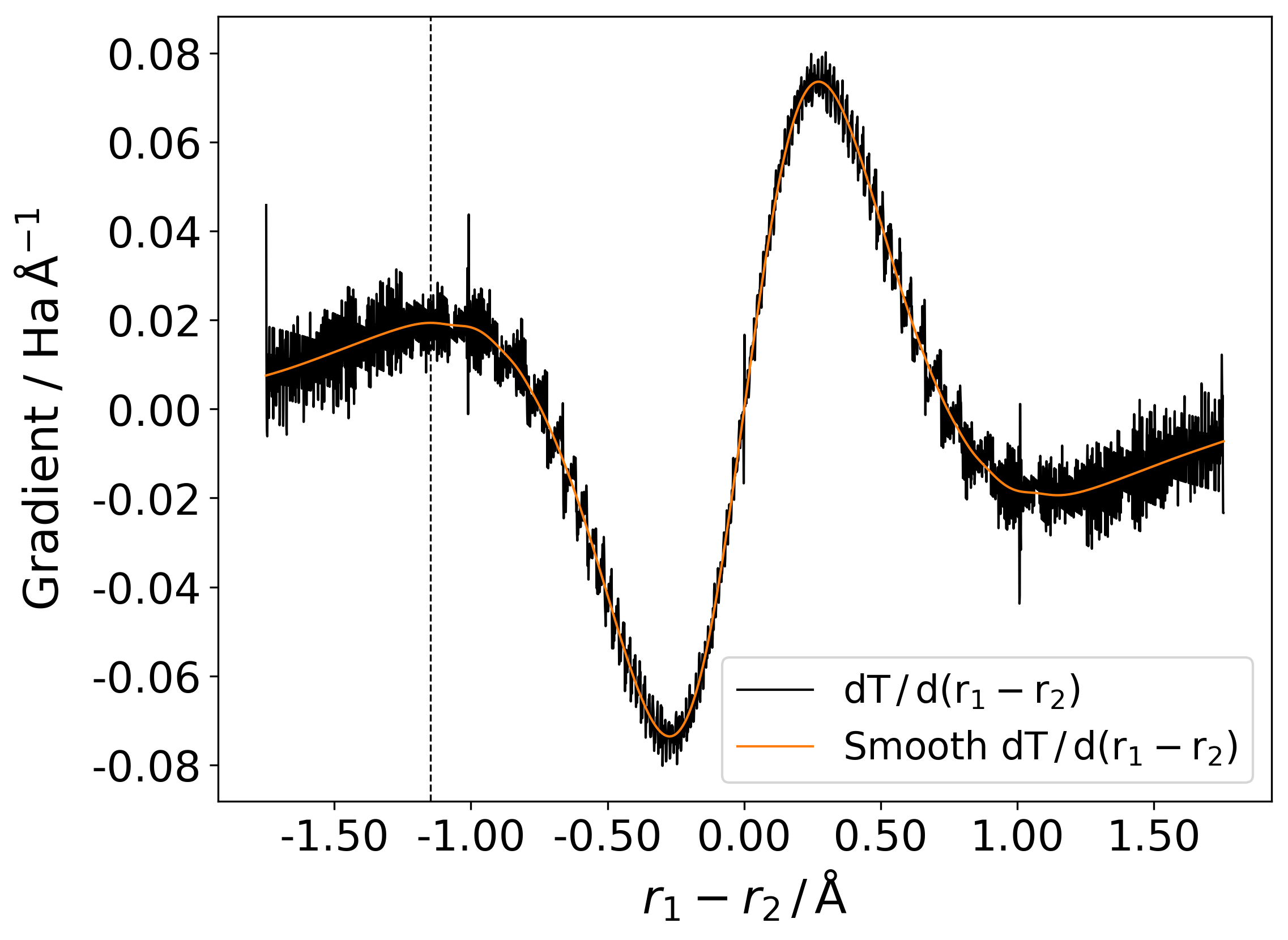}
\subcaption{}
\end{subfigure}
\caption{\textbf{(a)} CCSDT/aug-cc-pVTZ kinetic energy calculated from the numerical energy gradient and virial theorem, \textbf{(b)} gradient of the kinetic energy and the fitted spline}
\end{figure}

\section{Spin-specific charge transfer analysis of ALMO-EDA using non-perturbative COVP method}

Analysis of the electron density changes that occur through interactions between closed-shell and radical systems by Fukui et. al\cite{fukui1972} features noticeable charge transfer of $\alpha$ electrons (defining the excess spin as $alpha$). However, the density change due to $\alpha$-spin charge transfer is suggested to be almost exactly countered by the exchange interaction, which then leads to the conclusion that there is no net contribution from $\alpha$ electrons during bond formation. Further, the $\beta$ charge transfer from a closed-shell molecule to a radical is identified as the most important by Fukui et. al while $\alpha$ charge transfer in reverse direction is said to be the next most important contribution. 

Through our kinetic energy analysis, we have identified the role of $\alpha$ electrons in the existence of the \ce{H2} + \ce{H^.} transition state: $\beta$-electron delocalization counters the $\alpha$-electron pair-induced kinetic energy increase. This analysis is consistent with that of Fukui, since Pauli exclusion affects only $\alpha$ electrons in the \ce{H2} + \ce{H^.} system, and charge transfer by a $\beta$ electron is thought to drive the bond formation/breaking step. We emphasize that we have not isolated contributions that solely act to form the new bond as Fukui et. al did in their density analysis, so it is possible that the overall density contribution by the $\alpha$ electron charge transfer is canceled by the density change of exchange interaction in ALMO-EDA.

Interestingly, ALMO-EDA non-perturbative COVP\cite{nonpertCOVP} results show prominent energy stabilization through $\alpha$ electron charge transfer from the \ce{H^.} atom to the \ce{H2}. This effect is best described as the SOMO of \ce{H^.} interacting with the LUMO of \ce{H2}. Significant charge transfer also occurs between the $\beta$ electron of \ce{H2} to \ce{H^.}. In combination, this pair of charge transfer terms describes formation of the new covalent bond with concomitant cleavage of the existing bond. Notably, and consistently with both our analysis and that of Fukui, there is no energy stabilization associated with $\alpha$ electron delocalization from \ce{H2} to \ce{H^.}; this effect can be attributed to the Pauli exclusion principle.

\begin{figure}[H]
\centering
\includegraphics[width=0.8\textwidth]{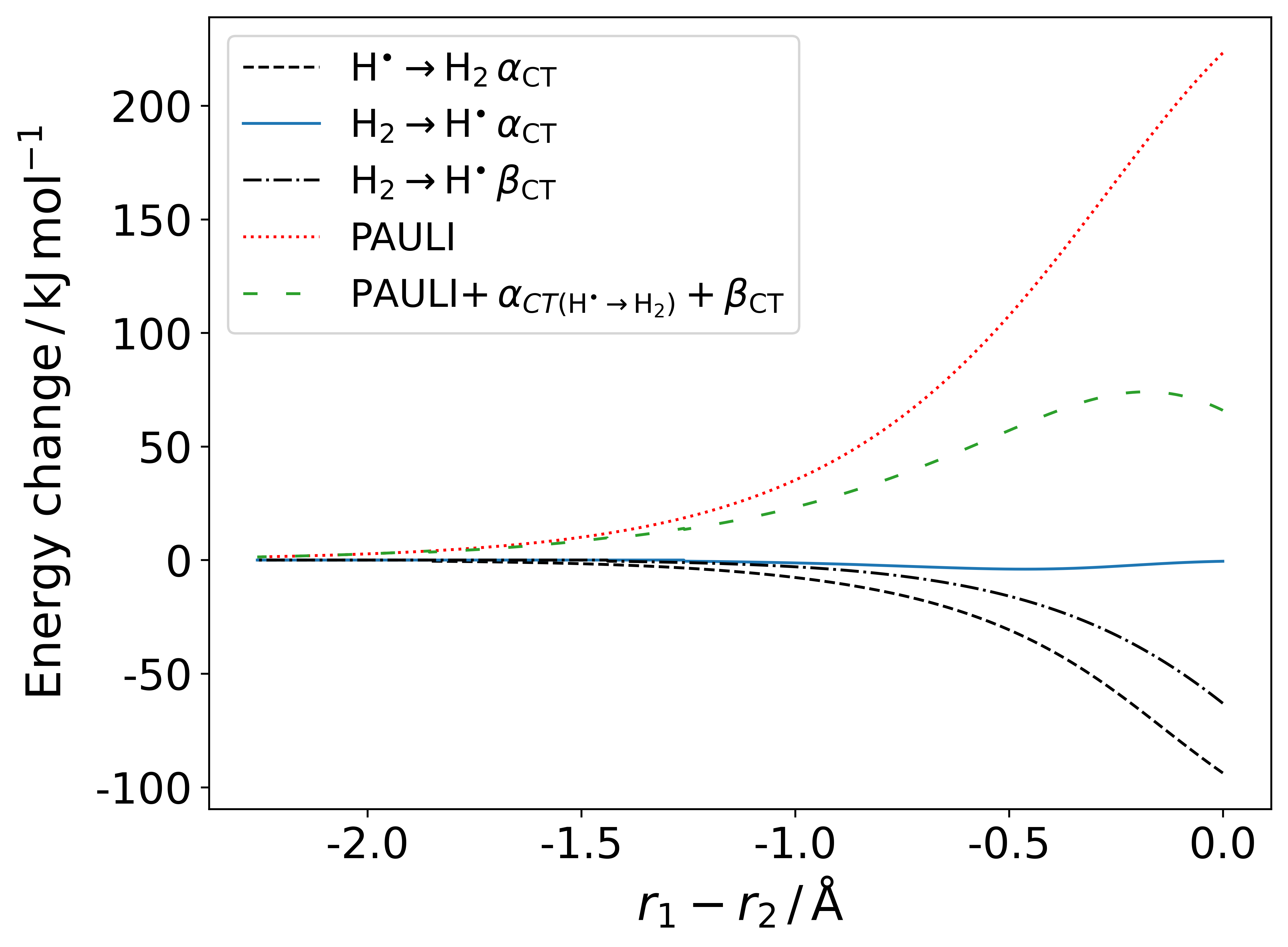}
\caption{Non-perturbative COVP charge transfer contributions of $\alpha$ and $\beta$ electrons overlaid with PAULI term}
\end{figure}

\bibliography{refs.bib}